\documentclass[11pt]{article}

\usepackage{epsfig,subfigure}
\usepackage{amssymb}
\usepackage[fleqn]{amsmath}
\usepackage{color}
\def\QED{\mbox{\rule[0pt]{1.5ex}{1.5ex}}}

\usepackage{graphicx}
\usepackage{color}

\usepackage{amssymb}
\usepackage{amsmath,amsfonts,amssymb}
\usepackage{verbatim}
\usepackage{stfloats}
\usepackage[bookmarks=false]{}

\newtheorem{theorem}{Theorem}

\newtheorem{lemma}{Lemma}

\newcommand\blfootnote[1]{%
  \begingroup
  \renewcommand\thefootnote{}\footnote{#1}%
  \addtocounter{footnote}{-1}%
  \endgroup
}

\begin{document}
\date{}
\title{Blind Interference Alignment for \\ Private Information Retrieval
}
\author{ \normalsize Hua Sun and Syed A. Jafar \\
}

\maketitle

\blfootnote{Hua Sun (email: huas2@uci.edu) and Syed A. Jafar (email: syed@uci.edu) are with the Center of Pervasive Communications and Computing (CPCC) in the Department of Electrical Engineering and Computer Science (EECS) at the University of California Irvine. }

\begin{abstract}
Blind interference alignment (BIA) refers to interference alignment schemes that are designed only based on channel coherence pattern knowledge at the transmitters (the ``blind'' transmitters do not know the exact channel values). Private information retrieval (PIR) refers to the problem where a user retrieves one out of $K$ messages from $N$ non-communicating databases (each holds all $K$ messages) without revealing anything about the identity of the desired message index to any individual database. In this paper, we identify an intriguing connection between PIR and BIA. Inspired by this connection, we characterize the information theoretic optimal download cost of PIR, when we have $K = 2$ messages and the number of databases, $N$, is arbitrary. 
\end{abstract}


\section{Introduction}

Interference alignment, in a variety of forms \cite{Jafar_FnT}, has  produced some of the most novel \cite{Cadambe_Jafar_int}, albeit  fragile \cite{Arash_Jafar_GC15, Arash_Jafar_IC} solutions for circumventing the interference barrier in wireless interference networks. Remarkably, the same ideas have found robust applications beyond wireless settings --- from  distributed storage  \cite{Cadambe_Jafar_Maleki_Ramchandran_Suh, Shah_IA} and network coding \cite{PBNA, Han_Wang_Shroff} to index coding  \cite{Maleki_Cadambe_Jafar, Sun_Jafar_nonshannon}, hat-guessing \cite{Riis_Hat} and topological interference management problems \cite{Jafar_TIM, Yi_Gesbert}. In this paper we report the discovery of another robust application of interference alignment outside the context of wireless networks. Specifically, we show how the idea of Blind Interference Alignment (BIA) that was originally introduced to exploit the diversity of coherence patterns in wireless interference networks, translates into new solutions for the Private Information Retrieval (PIR) problem previously studied by computer scientists. 

\subsection{Blind Interference Alignment}

The idea of blind interference alignment was introduced in \cite{Jafar_corr} to take advantage of the diversity of coherence intervals that  may arise in a wireless network. For instance, different channels may experience different coherence times and coherence bandwidths. A diversity of coherence patterns can also be artificially induced by the switching of reconfigurable antennas in pre-determined patterns. As one of the simplest examples of BIA, consider a $K$ user interference channel, where the desired channels have coherence time $1$, i.e., they change after every channel use, while the cross channels (which carry interference) have coherence time $2$, i.e., they remain unchanged over two channel uses. The transmitters are aware of the coherence times but otherwise have no knowledge of the channel coefficients. The BIA scheme operates over two consecutive channel uses. Over these two channel uses, each transmitter repeats its information symbol, and each receiver simply calculates the difference of its received signals. Since the transmitted symbols remain the same and the cross channels do not change, the difference of received signals from the two channel uses eliminates all interference terms. However, because the desired channels change, the desired information symbols survive the difference at each receiver. Thus, one desired information symbol is successfully sent for each message over 2 channel uses, free from interference, achieving $\frac{1}{2}$ DoF per message. Remarkably, with only the knowledge of channel coherence patterns and no knowledge of the actual values of channel coefficients, the BIA scheme is able to achieve the same DoF as possible with perfect knowledge of channel coefficients at all transmitters. Applications of BIA extend well beyond this simple example \cite{Wang_Gou_Jafar, BIA_Dynamic, BIA_IC, BIA_Z}. For instance, in the  $X$ channel comprised of $M$ transmitters and $K$ receivers, using only the knowledge of suitable channel coherence patterns, BIA schemes achieve $\frac{MK}{M+K-1}$  DoF, which cannot be improved upon even with perfect channel knowledge  \cite{Jafar_corr, Wang_Gou_Jafar}. 


\subsection{Private Information Retrieval}
In private information retrieval \cite{PIRfirst}, the goal is to (as efficiently as possible) allow a user to retrieve his desired information from data providers without disclosing anything about his interests. Note that the user can hide his interests trivially by requesting all the information available to the data providers, but that could be very inefficient (expensive). The goal of the PIR problem is to find the most efficient solution. Here is the problem description. We have $N \geq 2$ non-communicating databases\footnote{It turns out that the setting of $N=1$ is not interesting. It is proved in \cite{PIRfirst} that in order to be private in this case, the user has to download all the messages. Motivated by this result, \cite{PIRfirst} initiates the study of the benefit of non-communicating databases, and we adopt this model in this paper.}, each stores a set of $K \geq 2$ independent messages $W_1, \cdots, W_K$. A user wants one of the messages, say $W_i, i \in \{1,2, \cdots, K\}$, but requires each database to learn absolutely nothing (in the information theoretic sense)\footnote{There is another line of research, where privacy needs to be satisfied only for computationally bounded databases \cite{William, Yekhanin, CPIR}.}  about the retrieved message index, $i$. To do so, the user generates $N$ queries $Q_1, \cdots, Q_N$ and sends $Q_j, j \in \{1,2,\cdots, N\}$ to the $j$-th database. After receiving query $Q_j$, the $j$-th database returns an answering string $A_j$ to the user. The user must be able to obtain the desired message $W_i$ from all the answers $A_1, \cdots, A_N$. To be private, each query $Q_j$ and each answer $A_j$ must be independent of the desired message index, $i$. 

For example, suppose we have $N = 2$ databases and $K$ messages. To retrieve $W_1$ privately, the user first generates a random length-$K$ vector $[h_1, h_2, \cdots, h_K]$, where each element is uniformly distributed over a  finite field. Then the user sends $Q_1 = [h_1, h_2, \cdots, h_K]$ to the first database and $Q_2 = [h_1+1, h_2, \cdots, h_K]$ to the second database. Each database generates the answering string by taking a inner product of the query vector and the message symbols,
\begin{eqnarray*} \label{exeia}
\left[
\begin{array}{c}
A_1 \\
A_2
\end{array}
\right]
= \left[
\begin{array}{c}
h_1 \\
h_1 + 1
\end{array}
\right] W_1  +
h_2 \left[
\begin{array}{c}
1 \\
1
\end{array}
\right] W_2 + \cdots +
h_K \left[
\begin{array}{c}
1 \\
1
\end{array}
\right] W_K.
\end{eqnarray*}
The user can obtain $W_1$ by subtracting $A_1$ from $A_2$. Privacy is guaranteed as the query vectors $Q_1$, $Q_2$ are marginally random and independent of the desired message index, such that each database by itself can not tell which message is requested. The observant readers might have noticed the similarity between this  PIR scheme and the earlier example of BIA, where over two channel uses (across two databases), the channel (corresponding scalar in the query vector) for the desired message changes while the channels for the interfering messages remain the same. Note that the scalars in the query vector correspond to the channel coefficients, and the answering strings correspond to the received signal equations.
 
The cost of a PIR scheme is traditionally measured by the total amount of communication between the user and the databases, i.e., the sum of lengths of each query string (upload) and each answering string (download).
Prior work in theoretical computer science society studies the coding theoretic setting, where each message is one bit \cite{PIRfirst, YekhaninPhd, 2PIR}. 
In this paper, we consider the information theoretic formulation, where the size of each message goes to infinity. In this setting, the upload cost is negligible compared to the download cost\footnote{Similar observation is also made in \cite{Chan_Ho_Yamamoto}. 
The justification argument (traces back to Proposition 4.1.1 of \cite{PIRfirst}) is that the upload cost does not scale with the message size. This is because we can reuse the original query functions for each part of the message.}.
Therefore we  concentrate only on the download cost, measured in unit of the message size. For the above example, each message is one symbol and we download 2 equations in total, so that the download cost is 2.
We aim to characterize the optimal download cost, as an exact function of $N$ and $K$. 

\subsection{Overview of Contribution}
The main contribution of this paper is the discovery of an interesting connection between PIR and BIA such that a good BIA scheme  translates into a good PIR protocol. 
Let us see how to obtain a PIR protocol from a BIA scheme. The number of users in the BIA problem translates into the number of messages in the PIR problem. 
The received signals for user $i$  in BIA,  translate into the answering strings when message $W_i$ is the desired message in the PIR problem. The channel vectors associated with user $i$ in the BIA problem translate into the query vectors for desired message $W_i$ in the PIR problem. The privacy requirement of the PIR  scheme takes advantage of the observation that in BIA, over each channel use, the received signal at each receiver is statistically equivalent, because the transmitter does not know the channel values and the channel to each receiver has the same distribution. The most involved aspect of translating from BIA to PIR is that in BIA, the knowledge of the channel realizations across channel uses reveals the switching pattern, which in turn reveals the identity of the receiver. To remove this identifying feature of the BIA scheme, the channel uses are  divided into subgroups such that the knowledge of the switching pattern within each group reveals nothing about the identity of the receiver. Each sub-group of channel uses is then associated with a different database. Since the databases are not allowed to communicate with each other, and each sub-group of queries (channel uses) reveals nothing about the message (user), the resulting scheme guarantees privacy. Finally, the symmetric degrees of freedom (DoF) value per user in BIA is the ratio between the number of desired message symbols and the number of channel uses (received signal equations), and the download cost $\eta$ in PIR is the ratio between the total number of equations in all answering strings and the number of symbols of the desired message. In this way, the DoF value achieved with BIA translates into the reciprocal of the efficiency of the corresponding PIR protocol, i.e., $\eta = 1/\mbox{DoF}$. We summarize these connections in the following table.

\begin{table}[h]
\centering
\scalebox{1}{
\begin{tabular}{c | c}
   PIR & BIA \\
   \hline		
  Message & Receiver \\
  Queries &  Channel Coefficients \\
  Answers & Received Signals \\
  Efficiency & 1/DoF
\end{tabular}
}
\end{table}

As an application of the connection between BIA and PIR, we find the optimal efficiency $\eta$ for the PIR problem when we have $K = 2$ messages and the number of databases, $N$, is arbitrary.  We show that for this case, $\eta=1 + \frac{1}{N}$. Recall that with BIA the  $X$ channel with $N$ transmitters and $K=2$ receivers, achieves its optimal DoF value $1/(1+\frac{1}{N})=1/\eta$ \cite{Jafar_corr}. Indeed, for this setting the optimal PIR scheme for $K=2$ messages, $N$ databases, is found by translating from the BIA scheme for the $X$ channel with $K=2$ receivers and $N$ transmitters. A matching information theoretic outer bound is obtained to establish optimality. We note that the cost is strictly decreasing in $N$, meaning that having more databases strictly decreases the download cost, and when the number of databases becomes large, the cost approaches 1 (we just download what is needed). 
The best previously known scheme achieves the download cost of $1 + \frac{1}{N-1}$ \cite{Shah_Rashmi_Kannan}, which is strictly improved upon here. 

\section{System model}\label{sec:model}
We  define the system model for the PIR problem as follows. There are $K$ messages $W_1, \cdots, W_K$. We assume that the messages are independent and of the same size, i.e., 
\begin{eqnarray}
&& H(W_1, \cdots, W_K) = H(W_1) + \cdots + H(W_K), \label{h1}\\
&& H(W_1) = \cdots = H(W_K). \label{h2}
\end{eqnarray}
There are $N$ databases, and each database stores all the messages $W_1, \cdots, W_K$. A user wants to retrieve $W_i, i \in \{1, \cdots, K\}$ privately, i.e., without revealing anything about the message identity $i$ to any of the databases.

To retrieve $W_i$ privately, the user first generates $N$ queries $Q_1^{[i]}, \cdots, Q_N^{[i]}$, where the superscript denotes the desired message index. Each query $Q_j^{[i]}, j \in \{1,\cdots, N\}$ is a random variable with finite support. The queries are  independent of the messages, 
\begin{eqnarray}
I(W_1, \cdots, W_K; Q_1^{[i]}, \cdots, Q_N^{[i]}) = 0 \label{qwind}.
\end{eqnarray}
The user then sends query $Q_j^{[i]}$ to the $j$-th database. After receiving $Q_j^{[i]}$, the $j$-th database generates an answering string $A_j^{[i]}$, which is a deterministic function of $Q_j^{[i]}$ and the data stored (i.e., all messages $W_1, \cdots, W_K$), 
\begin{eqnarray}
H(A_j^{[i]} | Q_j^{[i]}, W_1, \cdots, W_K) = 0. \label{ansdet}
\end{eqnarray}
Each database returns to the user its answer $A_j^{[i]}$. From all answers $A_1^{[i]}, \cdots, A_N^{[i]}$, the user can decode the desired message $W_i$, 
\begin{eqnarray}
\mbox{[Correctness]} ~H(W_i | A_1^{[i]}, \cdots, A_N^{[i]}, {\color{black} Q_1^{[i]}, \cdots, Q_N^{[i]}}) = 0. \label{corr}
\end{eqnarray}

To satisfy the privacy constraint that each database learns nothing about the desired message index $i$ information theoretically, each query $Q_j^{[i]}$ must be distributed independently of $i$, 
\begin{eqnarray}
\mbox{[Privacy]} ~~~ I(Q_j^{[i]}; i) = 0, \forall j. \label{qi}
\end{eqnarray}
As the answering string is a deterministic function of the query and all messages, the answering string must be independent of $i$ as well, 
\begin{eqnarray}
I(A_j^{[i]}; i) = 0, \forall j. \label{ai}
\end{eqnarray}
{\color{black} For compact notation, we may sometimes ignore the desired message index superscript and simplify the notation as $Q_j/A_j$, when it would cause no confusion.}

The metric that we study in this paper is the download cost (efficiency) $\eta$. It is defined as the ratio between the total download cost and the size of the desired message, 
\begin{eqnarray}
\eta \triangleq \frac{\sum_{j=1}^N H(A_j^{[i]})}{H(W_i)}. \label{eta_def}
\end{eqnarray}
We aim to characterize the optimal (minimum) download cost $\eta^*$, over all private information retrieval schemes.

\section{Main Result}\label{sec:main}
We present our main result in the following theorem.

\begin{theorem}\label{thm:download}
For the private information retrieval problem with $2$ messages and $N \geq 2$ databases, the optimal download cost $\eta^*$ is $1 + \frac{1}{N}$.
\end{theorem}

Before presenting the proof for the general case of $N$ databases in Section \ref{sec:proof}, let us consider the $N=2$ setting.

\subsection{$N=2$ databases, $\eta^*=\frac{3}{2}$}\label{sec:dw1}


\subsubsection{Achievability}
The scheme follows from the BIA scheme of the $X$ channel with 2 transmitters and 2 receivers, where each transmitter has an independent message for each of the two receivers. The transmission takes place over 3 channel uses.  User 1 desires symbol $a_1$ from Transmitter 1 and symbol $a_2$ from Transmitter 2, and his channel changes from the first channel use to the second channel use and then remains the same from the second channel use to the third channel use, i.e., the channel pattern for User 1 is ${\bf h}^{[1]}(1), {\bf h}^{[1]}(2), {\bf h}^{[1]}(2)$. User 2 desires symbol $b_1$ from Transmitter 1 and symbol $b_2$ from Transmitter 2, and his channel remains the same from the first channel use to the second channel use  and then changes from the second channel use to the third channel use, i.e., the channel pattern for User 2 is ${\bf h}^{[2]}(1), {\bf h}^{[2]}(1), {\bf h}^{[2]}(2)$. The ${\bf h}^{[i]}(j)$ are random and statistically equivalent realizations of the $1\times 2$ channel vector from the two transmitters to Receiver $i$. In this channel, suppose the symbols are repeated by the transmitters over those channel uses where desired user's channel changes and the undesired user's channel remains the same, i.e., $a_1, a_2$ are repeated over the first two channel uses and $b_1,b_2$ are repeated over the second and third channel uses. 
Suppressing noise, the signal equations seen at the two receivers are
\begin{eqnarray}
\allowdisplaybreaks
{\bf y}^{[1]} = 
\left[ \begin{array}{c}
{\bf h}^{[1]}(1) \\
{\bf h}^{[1]}(2) \\
0
\end{array}
\right]
\left[ \begin{array}{c}
a_1 \\
a_2
\end{array}
\right] +
\left[ \begin{array}{c}
0 \\
{\bf h}^{[1]}(2) \\
{\bf h}^{[1]}(2)
\end{array}
\right]
\left[ \begin{array}{c}
b_1 \\
b_2
\end{array}
\right], \\
{\bf y}^{[2]} = 
\left[ \begin{array}{c}
{\bf h}^{[2]}(1) \\
{\bf h}^{[2]}(1) \\
0
\end{array}
\right]
\left[ \begin{array}{c}
a_1 \\
a_2
\end{array}
\right] +
\left[ \begin{array}{c}
0 \\
{\bf h}^{[2]}(1) \\
{\bf h}^{[2]}(2)
\end{array}
\right]
\left[ \begin{array}{c}
b_1 \\
b_2
\end{array}
\right].
\end{eqnarray}
Thus, each user sees two different linear combinations of desired symbols and only one linear combination (repeated) of interfering symbols. 
Subtracting the interference from its repeated observation allows each user access to two resolvable interference-free linear combinations of desired symbols. Thus, 2 symbols are successfully sent to each receiver over 3 channel uses, achieving $\frac{2}{3}$ DoF per receiver.

Now let us translate this BIA scheme into a PIR scheme, wherein $W_1 = \{a_1, a_2\}, W_2 = \{b_1, b_2\}$. To achieve download cost $\frac{3}{2}$, the total number of answering strings from the 2 databases must correspond to $3$ equations. The queries are the channels in the BIA problem. However, the channel uses must be grouped in such a way that the switching pattern does not reveal the identity of the receiver. For example, if the first two channel uses are in the same group then the identity of the receiver would be revealed (if the channel changes then it is Receiver 1 and if it does not then it is Receiver 2). However, if the first and third channel uses are placed in the same group, then we note that the channel changes within this group from one channel use to the next regardless of the receiver, thus revealing nothing about the identity of the  receiver. So this is the grouping we choose. Channel uses 1 and 3 are placed in one group and associated with Database 1, and channel use 2 is placed in the other group, associated with Database 2. Based on this grouping, the resulting queries $Q^{[i]}_j$ from the $j^{th}$ database when message $W_i$ is desired, are summarized in the following table.

\begin{table}[h]
\centering
\begin{tabular}{|c|c|c|}\hline
   $Q^{[i]}_j$ & Message  $W_i=W_1$ & Message: $W_i=W_2$ \\
   \hline		
Database: $j=1$&$ {\bf h}^{[1]}(1)$, ${\bf h}^{[1]}(2)$ &$ {\bf h}^{[2]}(1)$, ${\bf h}^{[2]}(2)$\\ \hline
Database: $j=2$&$ {\bf h}^{[1]}(2)$ &$ {\bf h}^{[2]}(1)$\\ \hline
\end{tabular}
\end{table}

The answering strings in the PIR problem are  the received signals of the BIA scheme. In correspondence with the queries, the answering string from the first database, $A_1$, is the received signal over the first and third channel uses and the answering string from the second database, $A_2$, is the received signal over the second channel use.
\begin{eqnarray}
A_1^{[1]} &=& 
\left[ \begin{array}{c}
{\bf h}^{[1]}(1)\\
0
\end{array}
\right]
\left[ \begin{array}{c}
a_1 \\
a_2
\end{array}
\right] +
\left[ \begin{array}{c}
0\\
{\bf h}^{[1]}(2) \\
\end{array}
\right]
\left[ \begin{array}{c}
b_1 \\
b_2
\end{array}
\right]
\\
A_1^{[2]} &=& 
\left[ \begin{array}{c}
{\bf h}^{[2]}(1)\\
0
\end{array}
\right]
\left[ \begin{array}{c}
a_1 \\
a_2
\end{array}
\right] +
\left[ \begin{array}{c}
0\\
{\bf h}^{[2]}(2) \\
\end{array}
\right]
\left[ \begin{array}{c}
b_1 \\
b_2
\end{array}
\right]
\\
A_2^{[1]} &=& 
\left[ \begin{array}{c}
{\bf h}^{[1]}(2) 
\end{array}
\right]
\left[ \begin{array}{c}
a_1 \\
a_2
\end{array}
\right] +
\left[ \begin{array}{c}
{\bf h}^{[1]}(2)
\end{array}
\right]
\left[ \begin{array}{c}
b_1 \\
b_2
\end{array}
\right] \\
A_2^{[2]} &=&
\left[ \begin{array}{c}
{\bf h}^{[2]}(1) 
\end{array}
\right]
\left[ \begin{array}{c}
a_1 \\
a_2
\end{array}
\right] +
\left[ \begin{array}{c}
{\bf h}^{[2]}(1)
\end{array}
\right]
\left[ \begin{array}{c}
b_1 \\
b_2
\end{array}
\right]
\end{eqnarray}
Clearly, from $A_1^{[1]}, A_2^{[1]}$, we can decode $W_1$, and from $A_1^{[2]}, A_2^{[2]}$, we can decode $W_2$, because the answering strings are the same as the received signals in the BIA scheme. 

Narrowing down to specific choices, several simplifications are readily noted. First, note that for the user to be able to recover his desired information, it suffices to set ${\bf h}^{[1]}(i)={\bf h}^{[2]}(i)={\bf h}(i)$, $i=1,2$ and ensure that ${\bf h}(1)$ is linearly independent of ${\bf h}(2)$. Note that this means the query to Database 1 is the same regardless of the message. Now we just need two linearly independent vectors, say $e_1=[1, 0]$ and $e_2=[0,1]$. In order to achieve privacy, it suffices if the user privately tosses a fair coin to decide if ${\bf h}(1)=e_1, {\bf h}(2)=e_2$ or ${\bf h}(1)=e_2, {\bf h}(2)=e_1$. 
The PIR scheme thus obtained can be specified explicitly as follows.

\begin{table}[h]
\centering
\begin{tabular}{|c|c|c|c|c|}\hline
&\multicolumn{2}{c}{Prob. 1/2} \vline&\multicolumn{2}{c}{Prob. 1/2} \vline\\ \cline{2-5}
    & Want $W_1 =(a_1,a_2)$ &Want $W_2=(b_1,b_2)$& Want $W_1=(a_1,a_2)$ &Want $W_2=(b_1,b_2)$ \\
   \hline		
Database $1$&$a_1$, $b_2$ &$a_1$, $b_2$& $a_2$, $b_1$&$a_2$, $b_1$\\ \hline
Database $2$& $a_2+b_2$ &$a_1+b_1$&$a_1+b_1$&$a_2+b_2$\\ \hline
\end{tabular}
\caption{The information that the user downloads from each database depending on his desired message and the outcome of a private fair coin toss.}
\end{table}
%
%
Thus, the user is equally likely to ask Database 1 for either ($a_1$, $b_2$) or ($a_2$, $b_1$), regardless of his desired message. Similarly, the user is equally likely to ask Database 2 for either $a_1+b_1$ or $a_2+b_2$, regardless of his desired message. Thus, from the perspective of each database, the information downloaded is independent of the desired message, guaranteeing privacy. Also note that in each case, the user has enough information to decode his desired message. This completes the achievability proof for $N=2$.


{\it Remark: We note that the PIR scheme presented above is asymmetric in the sense that we download $2$ equations from the first database and $1$ equation from the second database. This should not raise any concern, because without loss of generality, any PIR scheme can be made symmetric so that we download an equal number of equations from every database. To this end, we replicate the PIR scheme, so that now each message consists of 4 symbols, i.e., $W_1 = \{a_1, a_2, a_1', a_2'\}$, $W_2 = \{b_1, b_2, b'_1, b'_2\}$. For symbols $a_1, a_2, b_1, b_2$, we use the original PIR protocol. For symbols $a'_1, a'_2, b'_1, b'_2$ we also use the original PIR protocol (by generating another independent set of channels ${\bf h'}^{[1]}(1), {\bf h'}^{[1]}(2), {\bf h'}^{[2]}(1), {\bf h'}^{[2]}(2)$), but switch the roles of Database 1 and Database 2, so that for these symbols, we download $2$ equations from Database $2$ and one equation from Database 1. This produces a symmetric PIR scheme.}

\subsubsection{Converse}
Since the queries and answering strings for any given database are marginally independent of the desired message index, let us assume without loss of generality that $A^{[1]}_1=A^{[2]}_1=A_1, Q^{[1]}_1= Q^{[2]}_1=Q_1$, i.e., the query and answering string for the first database is the same, regardless of which message is requested. Then, from $A_1, A^{[1]}_2, A^{[2]}_2$ one can decode both messages, and from Fano's inequality we have,
\allowdisplaybreaks
\begin{eqnarray}
&& 2H(W_1) =  H(W_1,W_2| Q_1, Q^{[1]}_{2}, Q^{[2]}_{2})\\
&=&I(A_1, A^{[1]}_2, A^{[2]}_2; W_1, W_2| Q_1, Q^{[1]}_{2}, Q^{[2]}_{2}) \\
&=& H(A_1, A^{[1]}_2, A^{[2]}_2 | Q_1, Q^{[1]}_{2}, Q^{[2]}_{2}) \label{f1} \\
&=& H(A_1, A_2^{[2]} | Q_1, Q_2^{[1]}, Q_2^{[2]}) + H(A_2^{[1]} | A_1, A_2^{[2]}, Q_1, Q_2^{[1]}, Q_2^{[2]})  \\
&\leq& H(A_1, A_2^{[2]}) + H(A_2^{[1]} | A_1, A_2^{[2]}, Q_1, Q_2^{[1]}, Q_2^{[2]}, W_2) \label{f2} \\
&\leq& H(A_1) + H(A_2^{[2]}) + H(A_2^{[1]} | A_1, Q_1, Q_2^{[1]}, W_2) \label{f3}\\
&=& \eta H(W_1) + H(A_1, A_2^{[1]} |  Q_1, Q_2^{[1]}, W_2) - H(A_1 | Q_1, Q_2^{[1]}, W_2) \label{fx} \\
&=&  \eta H(W_1) + H(A_1, A_2^{[1]} |  Q_1, Q_2^{[1]}, W_2) \notag \\
&&~- 1/2 H(A_1 | Q_1, Q_2^{[1]}, W_2) - 1/2 H(A_2^{[1]} | Q_1, Q_2^{[1]}, W_2) \label{f5} \\
&\leq&  \eta H(W_1) + H(A_1, A_2^{[1]} |  Q_1, Q_2^{[1]}, W_2) \notag \\
&& - 1/2 H(A_1 | Q_1, Q_2^{[1]}, W_2) - 1/2 H(A_2^{[1]} | A_1, Q_1, Q_2^{[1]}, W_2) \notag \\
&=&  \eta H(W_1) + 1/2 H(A_1, A_2^{[1]} |  Q_1, Q_2^{[1]}, W_2) \\
&=&  \eta H(W_1) + 1/2 H(W_1) \label{fff}
\end{eqnarray}
where (\ref{f1}) is due to the fact that the answering strings are deterministic functions of the messages and queries, i.e., $H(A_1, A^{[1]}_2, A^{[2]}_2 | W_1, W_2, Q_1, Q^{[1]}_{2}, Q^{[2]}_{2}) = 0$, and (\ref{f2}) follows from the correctness condition that $W_2$ is a deterministic function of $A_1, A_2^{[2]}$. (\ref{f3}) follows from the property that dropping conditioning does not reduce entropy. In (\ref{fx}), we plug in the definition of $\eta$. In (\ref{f5}), we use a symmetric argument without loss of generality, so that $H(A_1 | Q_1, Q_2^{[1]}, W_2) = H(A_2^{[1]} | Q_1, Q_2^{[1]}, W_2)$ (this argument is proved in Lemma \ref{sym} in the next section). (\ref{fff}) follows from the fact that from $A_1, A_2^{[1]}$, we can decode $W_1$, and applying Fano's inequality. Rearranging (\ref{fff}), we have that $\eta \geq \frac{3}{2}$ and the outer bound proof is complete.


\section{Proof of Theorem \ref{thm:download}}\label{sec:proof}
\subsection{Outer Bound}
We define $A_{j:l}^{[i]} \triangleq \{A_{j}^{[i]}, A_{j+1}^{[i]}, \cdots, A_{l}^{[i]} \}, 1 \leq j \leq l \leq N, i \in \{1, 2\}$.
Similarly, $Q_{j:l}^{[i]} \triangleq \{Q_{j}^{[i]}, Q_{j+1}^{[i]}, \cdots, Q_{l}^{[i]} \}$.

Without loss of generality, we  assume that $A^{[1]}_1=A^{[2]}_1=A_1$, and that the PIR scheme is symmetrized by combining all permutations of any given PIR scheme as explained earlier. A consequence of this symmetry is formalized in the following lemma.

\begin{lemma}\label{sym}
[Symmetry]
Without loss of generality, we have
\begin{eqnarray}\label{syma}
&& H(A_1^{[1]}| Q_{1:N}^{[1]}, W_{2}) = \cdots = H(A_N^{[1]}| Q_{1:N}^{[1]}, W_{2}) \label{syma} \\
&& H(A_1^{[1]}| Q_{1:N}^{[1]}, W_{2}) \geq \frac{1}{N} H(W_1) \label{symaa}
\end{eqnarray}
\end{lemma}

{\it Proof:} We first prove (\ref{syma}). It follows from a permutation argument. Given a retrieval scheme described by a set of query and answer functions, we replicate the scheme $N!$ times and use one copy for each permutation of the databases. As a result, for the replicated scheme, the conditional entropy of the answer from each database is the same, and (\ref{syma}) follows.

Next we proceed to prove (\ref{symaa}). Note that from $A_{1:N}^{[1]}$, we can decode $W_1$. From Fano's inequality, we have
\begin{eqnarray}
 H(W_1) 
 = I(A_{1:N}^{[1]}; W_1 | Q_{1:N}^{[1]}, W_2)  = H(A_{1:N}^{[1]} | Q_{1:N}^{[1]}, W_2) \notag \\
\leq \sum_{j=1}^N H(A_{j}^{[1]}| Q_{1:N}^{[1]}, W_{2}) =  N H(A_{1}^{[1]}| Q_{1:N}^{[1]}, W_{2}) \notag
\end{eqnarray}
where the first line follows from the fact that the answers are deterministic functions of the messages and queries, and the second line is due to the property that dropping conditioning does not reduce entropy, and (\ref{syma}). Thus, (\ref{symaa}) is proved.
\hfill\QED

As in the $N=2$ case, from $A_1, A_{2:N}^{[1]}, A_{2:N}^{[2]}$, we can decode both messages $W_1, W_2$. 
From Fano's inequality, we have
\begin{eqnarray}
  2H(W_1) &=& H(W_1, W_2 | Q_{1:N}^{[1]}, Q_{2:N}^{[2]}) \\
&=& H(A_1, A_{2:N}^{[1]}, A_{2:N}^{[2]} | Q_{1:N}^{[1]}, Q_{2:N}^{[2]}) \label{oo1} \\
&\leq& H(A_1, A_{2:N}^{[2]})  + H(A_{2:N}^{[1]} | A_1, A_{2:N}^{[2]}, Q_{1:N}^{[1]}, Q_{2:N}^{[2]})  \\
&=&\eta H(W_1) 
+H(A_{2:N}^{[1]} | A_1, A_{2:N}^{[2]}, Q_{1:N}^{[1]}, Q_{2:N}^{[2]},W_2) \label{o2} \\
&\leq& \eta H(W_1) +H(A_{2:N}^{[1]} | A_1, Q_{1:N}^{[1]},W_2) \label{oo2} \\
&=&  \eta H(W_1)  + H(A_{1:N}^{[1]} | Q_{1:N}^{[1]}, W_{2}) - H(A_1 | Q_{1:N}^{[1]}, W_{2})  \\
&\leq&  \eta H(W_1) +  H(W_1) -  H(W_1)/N  \label{oo3}
\end{eqnarray}
where (\ref{oo1}) is due to the fact that the answering strings are deterministic functions of the messages and queries, (\ref{o2}) is due to the definition of $\eta$ and $W_2$ is a deterministic function of $A_1, A_{2:N}^{[2]}$, and in (\ref{oo3}) the second term follows from the fact that from $A_{1:N}^{[1]}$ we can decode $W_1$ and applying Fano's inequality and the last term is due to (\ref{symaa}). Rearranging (\ref{oo3}) produces the desired outer bound and the proof is complete.

An intuitive understanding of the outer bound proof is as follows. In total, we download $\eta H(W_1)$ amount of information, out of which $H(W_1)$ is the desired message and the remaining $(\eta - 1) H(W_1)$ is interference. When we request $W_2$, the interference would be $W_1$, and $(\eta - 1) H(W_1)$ would be the amount of the information about $W_1$ contained in the answering strings. From (\ref{symaa}), we know that this interference is at least $\frac{1}{N} H(W_1)$, so that $(\eta - 1) H(W_1) \geq \frac{1}{N} H(W_1)$, and $\eta \geq 1+ \frac{1}{N}$.

\subsection{Inner Bound}
The achievable scheme follows from the BIA scheme of the $X$ channel with $N$ transmitters and 2 receivers, which achieves $N/(N+1)$ DoF per receiver \cite{Jafar_corr}. 
The scheme operates over $N^2 -1$ channel uses. Transmitter $i \in \{1,\cdots, N\}$ wishes to send $N-1$ symbols $a_{1,i}, a_{2,i}, \cdots, a_{N-1,i}$ to User 1 and another $N-1$ symbols $b_{1,i}, b_{2,i}, \cdots, b_{N-1,i}$ to User 2.
The channel coherence pattern is given by the following grid, where ${\bf h}^{[i]}(j)$ are random and statistically equivalent realizations of the $1 \times N$ channel vector from the $N$ transmitters to Receiver $i$. 
\begin{table}[h]
\centering
\scalebox{1}{
\begin{tabular}{|c|c|c|c|c|c|}
   \hline		
   ${\bf h}^{[2]}(1)$ & 1 & 2 & $\cdots$ & $N-1$ & $N$ \\
   \hline		
   ${\bf h}^{[2]}(2)$ & $N+1$ & $N+2$ & $\cdots$ & $2N-1$ & $2N$ \\
     \hline		
  $\vdots$ & $\vdots$ & $\vdots$ & $\vdots$ & $\vdots$ & $\vdots$ \\
     \hline
    ${\bf h}^{[2]}(N-1)$  & $N^2 - 2N + 1$  & $\cdots$ & $\cdots$ & $N^2 - N-1$ & $N^2 - N$ \\
   \hline		
  ${\bf h}^{[2]}(N)$  & $N^2 - N + 1$ & $\cdots$ & $\cdots$ & $N^2-1$ & \\
     \hline		
       & ${\bf h}^{[1]}(1)$ & ${\bf h}^{[1]}(2)$ & $\cdots$ & ${\bf h}^{[1]}(N-1)$ & ${\bf h}^{[1]}(N)$ \\
       \hline
\end{tabular}
}
\end{table}

In this channel, suppose symbols are repeated by the transmitters over channel uses where desired users' channel changes and the undesired users' channel remains the same, i.e., $a_{p, i}, p \in \{1, \cdots, N-1\}$ are repeated by Transmitter $i$ over the channel uses where the channel to User $2$ remains ${\bf h}^{[2]}(p)$, and $b_{q,i}, q \in \{1, \cdots, N-1\}$ are repeated by Transmitter $i$ over the channel uses where the channel to User 1 remains ${\bf h}^{[1]}(q)$. Thus, the received signal at each receiver has the property that  the desired signal has rank $N^2 - N$ and the interference has rank $N-1$ \cite{Jafar_corr}. 
Further, the desired signal does not overlap with the interference, and each user is able to achieve DoF $N/(N+1)$.

Now let us translate the BIA scheme to a PIR scheme, wherein each message consists of $N^2 -N$ symbols. To achieve download cost $\frac{N+1}{N} = \frac{N^2 - 1}{H(W_1)}$, the answering strings from all $N$ databases must consist of $N^2 - 1$ equations. The queries are the channels in the BIA problem. To ensure that the switching pattern does not reveal the identity of the receiver, the channel uses are grouped as follows.
Channel uses at the diagonal of the grid, i.e., channel uses $1, N+2, 2N+3, \cdots, N^2 - N-1$ ($N-1$ in total), are placed in the first group and associated with Database 1. Channel uses at the diagonal of the grid shifted cyclicly to the right by 1, i.e., channel uses $2, N+3, \cdots, N^2 - N, N^2 - N + 1$ ($N$ in total), are placed in the second group and associated with Database 2.  Channel uses at the diagonal of the grid shifted cyclicly to the right by 2 are placed in the third group, associated with Database 3, and so on.  
As a result, the channel changes within this group from one channel use to the next regardless of the receiver, thus revealing nothing about the identity of the receiver. 
When message $W_i$ is desired, the query vector sent to the $j^{th}$ database is comprised of the channels to Receiver $i$ in the $j^{th}$ group. 
The answering string from each database is  comprised of the received signal equations corresponding to the channel uses in each query vector. When message $W_i$ is desired, the answering string is the received signal at the Receiver $i$. As the answering strings are the same as the received signal in the BIA scheme, we can decode the desired message.

Similar to the proof for $N = 2$ presented in the previous section, let us specify the choices. To ensure decodability, we set ${\bf h}^{[1]}(i)={\bf h}^{[2]}(i)={\bf h}(i)$, $i=1,\cdots, N$ and ensure ${\bf h}(i)$ are linearly independent. For such a purpose, we just need $N$ vectors $e_0, \cdots, e_{N-1}$, where $e_{i-1}$ is an $N \times 1$ vector, where the $i^{th}$ entry is one and all other entries are zero. To be private, it suffices if the user privately draw a number $l$ uniformly from the set $\{0, 1, \cdots, N-1\}$ and set ${\bf h}(i) = e_{i-1 + l \mod N}$.
This completes the achievability proof.

\section{Discussion} \label{sec:disc}
We identified an interesting connection between PIR and BIA, which ensures that a good BIA scheme can be translated into a good PIR protocol. Based on this insight we found the optimal download cost for the $K = 2$ messages setting with $N\geq 2$ databases.

An immediate open problem is to find the optimal download cost when $K \geq 3$. Here the best known scheme achieves the cost of $1+ \frac{1}{N-1}$ \cite{Shah_Rashmi_Kannan}. Interestingly, the scheme proposed in \cite{Shah_Rashmi_Kannan} can also be translated from a BIA scheme. Consider a $K$ user MISO interference channel, where each transmitter is equipped with $N-1$ antennas and each receiver is equipped with a single antenna. The BIA scheme achieves $1/(1+ \frac{1}{N-1}) = \frac{N-1}{N}$ DoF per receiver, where each transmitter sends $N-1$ symbols to its desired receiver over $N$ channel uses. Over these $N$ channel uses, the desired channel changes after every channel use, the cross channels remain unchanged, and each transmitter repeats its information symbols (one from each antenna). Thus, each receiver sees only one repeated linear combination of interfering symbols and he can project the received signal equations along the direction of the nullspace of the all 1 vector to extract the $N-1$ desired symbols. Following the general connection between BIA and PIR, this BIA scheme can be easily translated into a PIR scheme, where each message consists of $N-1$ symbols and we download one equation from each of the $N$ databases. Each channel use in BIA is associated with one database in PIR, so that when $W_i, i \in \{1,\cdots, K\}$ is desired, the query to the $j^{th}, j \in \{1,\cdots, N\}$ database in PIR is the channel vector from all transmitters to the $i^{th}$ receiver over the $j^{th}$ channel use in BIA and the answering string from the $j^{th}$ database in PIR is the received signal equation at the $i^{th}$ receiver over the $j^{th}$ channel use in BIA. This completes the description of the PIR scheme with download cost $1 + \frac{1}{N-1}$. However, the cost of $1+ \frac{1}{N-1}$ does not match the converse provided by direct extensions of the techniques introduced in this paper. This leaves open the possibility of another BIA scheme that may provide a better efficiency. Notably, in the regime where the number of messages becomes large, i.e., $K \rightarrow \infty$, the download cost of $1+ \frac{1}{N-1}$ is asymptotically optimal. The converse can be established as follows, which is a simple extension of our outer bound proof. Without loss of generality, we assume $A_1^{[1]} = \cdots = A_1^{[K]} = A_1$. Then, from $A_1, A_{2:N}^{[2]}, \cdots, A_{2:N}^{[K]}$ one can decode all $K$ messages $W_1, \cdots, W_K$, so that the entropy of $A_1, A_{2:N}^{[2]}, \cdots, A_{2:N}^{[K]}$ is no less than $KH(W_1)$. By a simple extension of Lemma \ref{sym} to $K$ messages setting, we know that by symmetry, the entropy of each term in $A_1, A_{2:N}^{[2]}, \cdots, A_{2:N}^{[K]}$ is the same and is equal to $\eta H(W_1)/N$. As there have $1 + (K-1)(N-1)$ terms in total, we have 
\begin{eqnarray*}
\eta \geq \frac{NK}{1 + (K-1)(N-1)} = \frac{N}{\frac{1}{K} + (1- \frac{1}{K})(N-1)}
\end{eqnarray*}
and when $K \rightarrow \infty$, $\eta \geq 1 + 1/(N-1)$.

\bibliographystyle{IEEEtran}
\bibliography{Thesis}
\end{document}